# Tuning near field radiative heat flux through surface excitations with a metal insulator transition

P.J. van Zwol, L. Ranno, J. Chevrier

*Institut Néel, CNRS and Universite Joseph Fourier Grenoble, BP 166 38042*
*Grenoble Cedex 9, France*

The control of heat flow is a formidable challenge due to lack of good thermal insulators. Promising new opportunities for heat flow control were recently theoretically discovered for radiative heat flow in near field, where large heat flow contrasts may be achieved by tuning electronic excitations on surfaces. Here we show experimentally that the phase transition of $VO_2$ entails a change of surface polariton states that significantly affects radiative heat transfer in near field. In all cases the Derjaguin approximation correctly predicted radiative heat transfer in near field, but it underestimated the farfield limit. Our results indicate that heat flow contrasts can be realized in near field that can be larger than those obtained in farfield.

It is said that when Planck postulated his law on blackbody radiation, he realized that his theory was not valid when the distance between radiating blackbodies would be smaller than the peak wavelength or about 10 micron. Indeed when this regime is entered radiative heat transfer (RHT) is much increased [1]. A theoretical framework was developed to account for near field (NF) effects [1] and first experimental evidence for enhancement beyond the farfield (FF) limit was found in the seventies of the last century [2] but it was not until this millennium that precise verification became possible with the advent of modern precision measurements based on Scanning Tunneling Microscopy (STM) [3] or Atomic Force Microscopy (AFM) [4,5] and later macroscopic plates [6,7].

It is now known that surface excitations can enhance NF RHT by several orders of magnitude [8,9], specifically when they have a wavelength near the maximum of the Planck blackbody spectrum. Controlling NF RHT by means of *'surface excitation tuning'* has therefore recently attracted increased theoretical interest. Propositions include RHT control; with temperature [10], anisotropy [11], by tuning plasmons in graphene [12], or doped silicon [13], thinfilms [14], roughness [15] or by employing phase change materials [16], allowing unprecedented control of heat flux [17]. A theoretical work in [16] showed that the metal-insulator transition (MIT) of $VO_2$ at 68ºC involves a change in phonon-polariton states that can be used to switch NF RHT by orders of magnitude. Here we experimentally investigate this system and show that NF RHT can be in situ and repeatedly modified.

To measure RHT we employ a setup that has been described in length in refs [5,18]. Our methods are not different from those in [5]. However we do stress that specifically for the analysis of FF RHT we use the method described in ref [18] in which issues are considered that were not treated in ref [5].

Briefly we employ a room temperature high vacuum interferometric AFM ($5·10^{-8}$ mBar). As probe we use a 320μm bilayer microlever (*Veeco MLCT-O10*) to which a *40 micron* diameter sodalime glass sphere is attached at its free end (fig. 1). Such probes bend in response to heat flux [4, 19]. This lever is at room temperature under the experimental conditions [18] and resides vertically above a sample that can be heated to minimize force contributions. In order to calibrate our system we use two different glass samples, one being a 0.2mm thick *fused quartz* sample (*Electron Microscopy Sciences*) and the other a *500nm $SiO_2$ thinfilm* grown on Si (fig. 1c). We can calibrate our system both to NF [5] and FF RHT [18].

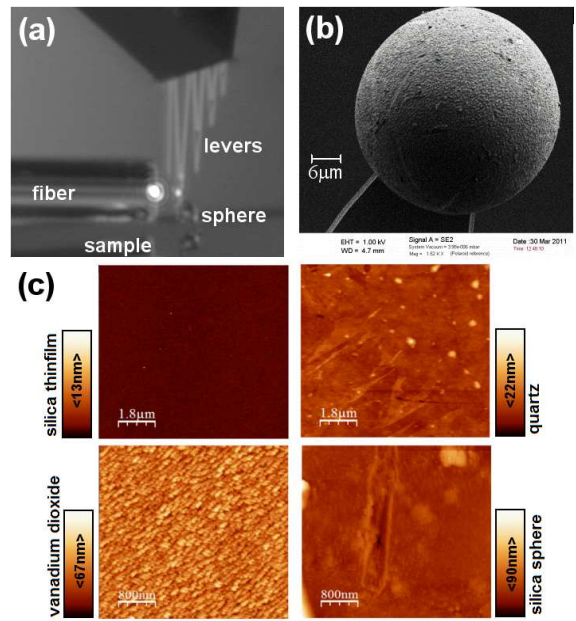

**Figure 1:** *(a) Optical image of the fiber, the lever with sphere and the sample with a visible reflection of the sphere and the fiber. (b) SEM image of the sphere. (c) Scans of the involved surfaces. For the sphere the peak height varies from place to place and can be as high as 200nm.*

Regardless of the fact that our lever is rotated 90 degrees to minimize forces (fig. 1a), we measured contri-butions of electrostatic forces and potentially dispersion forces for all investigated samples which we attribute to a torque effect as we could not minimize them by rotating the sample (fig 2a inset). We found that the electrostatics only affected NF RHT measurements, as FF RHT was found to be constant within 3% in the range 40-200micron [18]. For quartz and $VO_2$ the measured electrostatic forces were small but we measured particularly strong forces for the thermally grown glass thin film (fig. 2a). Such forces are due to uncontrolled charges in or on the surface of the material. These forces are measured when no heat is applied *(ΔT=0)* and are subtracted from the curves measured at ΔT>0 such as in fig. 2d. Note that while heating the plate, the NF RHT leads to lever motion opposite to motion due to electrostatic forces (fig 2a,d).

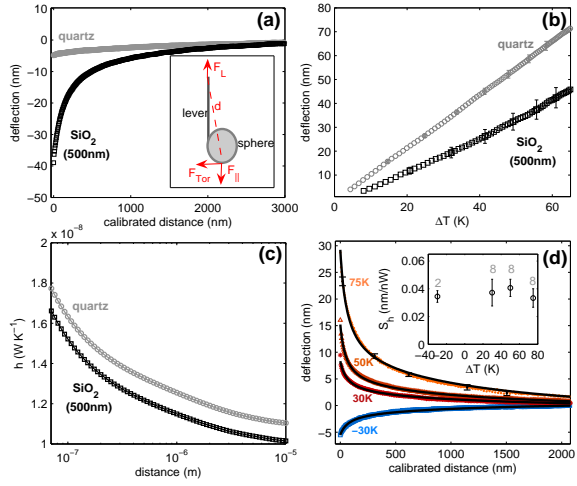
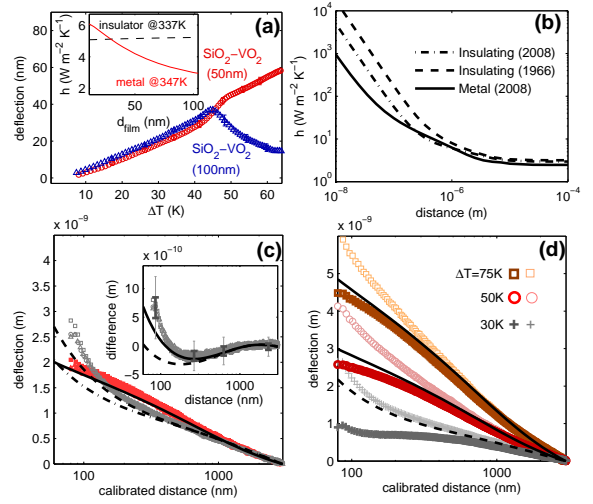

*Figure 2:* *(a)* Measured deflection-distance curves (piezo calibration is applied) for the two glass samples at ΔT=0K. They are attributed to a torque d×$F_{//}$, which results in $F_{Tor}$ because $F_L$ and $F_{//}$ do not coincide at the lever axis (1a inset). *(b)* FF curves (at 40μm) RHT as function of ΔT, the standard deviation denotes variation from place to place on the surface *(c)* NF RHT theory in the plate-sphere setup for a sphere diameter of 40μm. *(d)* Deflection-distance curves for the $SiO_2$ thin film sample at different ΔT, from which the electrostatic contribution in fig. 2a at ΔT=0K is subtracted. The curves are fitted to theory (lines) to calibrate the contact point $d_0$ and lever sensitivity $S_h$. Each curve is an average of six curves taken at the same place where the standard deviation is shown for the ΔT=75K curve. The inset in (d) shows values and standard deviation of $S_h$ as obtained for different places (number of places are indicated for each ΔT)

*Figure 3* *(a)* Measured FF RHT versus ΔT for the 50 and 100nm $VO_2$ films on a Sapphire substrate. The inset shows theory in FF versus $VO_2$ film thickness *(b)* Parallel plane theory of NF RHT at 300K between $SiO_2$ and $VO_2$ (50nm on Sapphire). Results for two sets of dielectric data for insulating $VO_2$ are shown. *(c)* Three averages of RHT measurements done at 9 different places on the surface are shown for both $VO_2$ phases (ΔT=30K minus ΔT=0K, open symbols, ΔT=75K minus ΔT=50K, closed symbols). Theory is for the metallic (─) and insulating (--) phase (see legend fig. 3b). The inset shows differences between the metal and insulating curves. Thin error bars depict place dependent variation in the measurements, thick error bars are standard deviations of the averages. *(d)* Averages of all 27 measurements obtained at different places, and theory (see legend 3b, 2 curves for metallic $VO_2$ are shown). Fat symbols mean ΔT=0K curve is not subtracted, thin symbols mean that the ΔT=0K curve is subtracted.

We repeat our RHT measurement at several different places on both glass surfaces and at different temperatures in the range *(ΔT=-30,75K)*. Regardless of the strength of the electrostatic contribution we measured the same NF RHT increase for both the quartz and the glass thin film sample. Our results fitted well to standard RHT theory [9] (fig. 2c) using dielectric data for silica [16] and the Derjaguin approximation [20]. We deduce the cantilever sensitivity $S_h$= $deflection_{lever}/RHT_{theory}$ (in nm/nW) and the point of contact $d_0$ from these fits (fig. 2d) [5] and found $S_h$=0.039±0.009nm/nW and $d_0$=83±21nm for fused quartz and $S_h$=0.037±0.008nm/nW and $d_0$=66±9nm for the $SiO_2$ thin film. The standard deviations are not obtained from the fit such as in ref. [5] but are experimental ones obtained from multiple measurements at different temperatures and places. The difference and variation of the fitted contact points corresponded to the roughness statistics of our samples and sphere (fig. 1b,c). Thus the measured NF RHT increase is the same for the two cases within experimental uncertainties. This is also theoretically reproduced and can be physically understood as based on interacting phonon polaritons.

When we compare the FF RHT measurements (fig. 2b) to the theoretical value for pure silica surfaces as predicted by the Derjaguin approximation (8nW/K, with T=300 and 360K for the two plates) we obtain $S_h$=0.125nm/nW. While NF measurement are only sensitive to the sphere, in FF we measured that the lever contributes about 25% of the total RHT as shown in ref. [18]. As such we obtain $S_h$=0.094nm/nW for the sphere only, which is still over a factor two too large. This factor two is largely explained by applying corrections from exact theory [21], which indicates that the Derjaguin approximation underestimates FF RHT by about 45% for our sphere yielding $S_h=0.052nm/nW$, which is close to the measured value in NF.

We believe that the remaining difference may be explained by variations from place to place on the surface (see fig. 2b), or by errors due to repositioning the fiber [18]. These effects lead to measured variations in RHT which are at the 10% level. Typical variations in emissivity values reported for glass surfaces may also play a role. Our measurements show that for the glass thin film FF RHT is reduced by 35% as compared to the bulk glass plate (fig. 2b). This decrease is reproduced by RHT theory for thin films using multilayer reflection coefficients, for 500nm silica on silicon, which predicts an 18% decrease yielding $S_h=0.041nm/nW$. Thus we find good agreement between experiment and theory in NF and FF for two different samples, which defines the calibration in the analysis for the results on $VO_2$ presented below.

Earlier we have reported on FF RHT measurements for a 100nm $VO_2$ film grown on sapphire and found a factor of five decrease when the MIT occurred [18]. This film had however *12nm RMS* roughness deeming it unsuitable for NF RHT measurements. Thus we have grown another $VO_2$ film with thickness 50nm to reduce the roughness to *5nm RMS*. Both films exhibited a three orders of magnitude or more change in conductivity at the MIT. The FF behavior of the 50nm $VO_2$ film however completely changed in the metallic state, exhibiting instead of a factor five reduction, an increase of 20% in the FF RHT at the MIT (fig. 3a). This decrease of FF RHT with film thickness was also theoretically observed. But the effect was not as strong (fig. 3a inset) and for film thickness 50nm we measure a 20% increase upon MIT (fig. 3a), while theory predicts a 20% decrease (fig. 3b). This, we believe, is due to a lack of knowledge of exact dielectric properties of our samples and possibly surface morphology. We found for example that lowering the Drude absorption for metallic $VO_2$ by 30% yielded better agreement with theory to within 20%. The quality of our sample may be somewhat different from the ones whose optical properties are used in [16]. For example the conductivity depends strongly on the preparation conditions even for gold (see ref. [16] and references therein).

For two parallel plates, theoretically a contrast of up to a factor of 100 was predicted [16] for NF RHT between a glass surface and bulk $VO_2$ that undergoes a MIT. For $VO_2$ as a thin film the RHT is influenced by the sapphire substrate specifically for the metallic phase, reducing contrast to about a factor of 20 (fig. 3b). In our measurement the largest contrast decrease is however due to the use of a plate-sphere setup (to avoid parallelism problems [6,7]). For this case the contrasts are reduced to about 50% for the current attainable distance range due to RHT contributions from distance range above 1μm where the RHT contrast between the two phases is reversed (fig. 3c). For simplicity reasons for the theory we ignore dielectric anisotropy for the $VO_2$ film [11, 16] and the sapphire substrate [22] in our calculations. We use multilayer reflection coefficients where necessary. Frustrated modes for anisotropic materials may increase the heat flux somewhat by up to 50% in a plate-plate configuration [11] which becomes lower in our plane-sphere case. Surface roughness is not treated but may give a contribution in the order of 10% at the smallest probed distances [15], and we use the Derjaguin approximation to obtain RHT results in the plane-sphere configuration [23]. We do perform the calculations for both the 1966 dielectric data for $VO_2$ of Barker [24] and the 2008 set of data for Qazilbatch et al [25] for the insulating case. The calculations in the insulating state are much more sensitive to variations in measured dielectric data than the metallic case which has simple Drude behavior. Instead the calculations for the insulating phase vary due to strong dependence on the frequencies and strengths of the involved phonon-polaritons.

We repeated NF RHT measurements at 27 different places on the $VO_2$ surface for temperature differences $\Delta T=0, 30, 50$ and $75K$. For every place and $\Delta T$ we took six curves. Note that the phase transition happens at around $\Delta T=43K$. As with glass we subtracted the electrostatic contribution i.e. the $\Delta T=0K$ curve from the $\Delta T=30K$ curve for the insulating phase. The electrostatic force may change at the MIT, however it is not possible to measure the electrostatic force independently from RHT for $\Delta T>0$ as RHT is measured at the same time. In order to compare the pure metallic phase to the pure insulating phase we chose to subtract the electrostatic contribution for the metallic phase by subtracting the $\Delta T=50K$ curve from the $\Delta T=75K$ curve. We compare the resulting curve to the one resulting from the $\Delta T=30K$ curve minus the $\Delta T=0K$ curve. We also calculated that the NF RHT would be nearly identical for these two curves to within a few percent if $VO_2$ would not undergo a phase transition.

To compare measurements and theory we used the calibration values as obtained from the silica-silica measurements (fig. 2d) i.e. $S_h=0.037nm/nW$ and $d_0=80nm$. Because the FF RHT contribution was large in our plane-sphere setup as compared to the NF RHT increase (in the attainable distance range) we decided to show only the measured NF RHT increase below 3μm (where we start the RHT-distance curves). At distance 3μm, the RHT is shifted vertically to zero for both theory and measurement, and for both phases. Then the measurements are shifted horizontally to $d_0$ and theory is scaled with $S_h$. Thus in this comparison between measurement and theory there are no parameters adjusted in figs. 3c,d. To obtain the measured RHT one can just multiply the data in figs 3c with $S_h$ and add appropriately the FF contribution using fig 3a.

In total 27 measurements were done at different places on the surface and 3 averages of 9 such measure-ments are shown fig. 3c (see also supplemental material). Below 200nm a strong increase of RHT was found for the insulating state. This is associated to the phonon-polariton contribution which leads to order of magnitude RHT contrasts upon MIT for the small part of the sphere that is closest to the surface fig. 3b. To better reveal the measured difference in NF RHT between the metallic and insulating phases we subtracted the metal from the insulator case from fig. 3c. The result is shown in the inset of fig. 3c. Both measurements and theory as obtained with the two sets of dielectric data revealed the same behavior yielding a nontrivial difference curve.

At last we discuss the observed electrostatic forces. For glass surfaces we found no compelling evidence for a temperature dependent electrostatic force in the sense that the RHT theory fitted well at all temperatures yielding a temperature independent $S_h$ (fig. 2d inset). Furthermore we found the same $S_h$ for both glass surfaces regardless of the large measured difference in electrostatic force at $\Delta T=0K$. At last $S_h$ in FF is not affected by the electrostatic force and was found to be similar to that in NF. For $VO_2$ in the metallic phase we found that theory and experiment at $\Delta T=50$ and $75K$ agree with very similar differences (residuals) between theory and experiment (fig. 3d), indicating no strong temperature dependent effect from forces or other parasitic signals. In the metallic phase the theory agrees better with experiment when the $\Delta T=0K$ curve is not subtracted from the $\Delta T=50K$ and $\Delta T=75K$ curves, while for the insulating phase the inverse is the case (fig. 3d). This suggests that the electrostatic force is smaller in the metallic phase. Thus fig. 3d is here to underline the necessity to subtract the $\Delta T=50K$ curve from the $\Delta T=75K$ curve in the metallic phase in fig. 3c as described above.

Concluding, we have shown experimentally that NF RHT is enhanced by the change in phonon-polariton states during the metal-insulator transition of $VO_2$. This reveals that very large heat flow contrasts in NF are present for the area of the sphere closest to the plane, as compared to FF RHT or bulk heat conduction [16]. We believe that our measurements are an important step toward controllable infrared photonic devices, as besides in situ temperature based control of the MIT of $VO_2$, also electric or photonic control is possible. This may yield a new class of switchable NF heat flow devices [17, 26, 27].

**Acknowledgements:** We gratefully acknowledge support of the Agence Nationale de la Recherche through the Source-TPV project ANR 2010 BLAN 0928 01. We thank K. Joulain and S. Thiele for help on theory and electrical characterisation of $VO_2$.

# Supplemental

## Near field RHT measurements and averaging procedure

In figure 1 of this supplemental we show the raw measurements of near field radiative heat transfer (RHT) as done at 9 different places. At each place six curves are taken for each temperature ΔT= 0K, 30K, 50K and 75K. These six curves are averaged to reduce noise. Then the resulting ΔT= 0K curve is subtracted from the ΔT=30K curve. The ΔT=50K curve is subtracted from the ΔT=75K curve. In this way the electrostatic force is properly subtracted from the RHT in both phases even if the electrostatic force changes at the MIT. In 9 out of 10 curves, for the distances <150nm, the insulating phase always has highest RHT, which is consistent with the presence of phonon coupling. For distances >150nm, for most curves the metallic phase has the highest RHT. The 9 curves below are a representative set for our measurements. Nine of these curves are used for 1 average curve in figure 2.

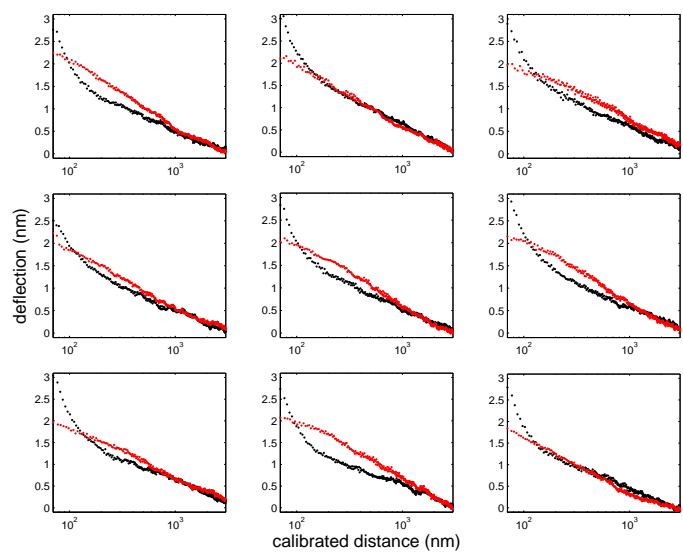

*Figure 1. Raw data of measurements of near field RHT versus distance between a silica sphere (40micron) and a $VO_2$ plate. Data is taken at 9 different places. Red curves are taken in the metallic phase and depict the ΔT= 75K curve from which the ΔT=50K curve is subtracted. Black data is for the insulating phase, and depict the ΔT= 30K curve from which the ΔT=0K curve is subtracted.*

In figure 2(a) the error bars due to spatial variation of the measured RHT on the surface overlap. However first of all this error does not do justice to the data in fig. 1, from which it is evident that the general behavior in the curves follows that of the average. Furthermore fig. 3 indicates that the spatial variation in measured RHT is random, hence the errors are added up squared when subtracting the metallic phase from the insulating phase to produce the difference curve in fig. 2b. It is seen that the shape of the difference curve in fig.2b has significant statistical meaning even if spatial variation is considered. However in fig. 2b three averages of 9 curves are shown to reduce random errors. Thus it is more reasonable to show the standard deviation of the average. If once again random errors are assumed this error is reduced by $\sqrt{9}=3$. It can be seen that all averages are within the resulting error bars.

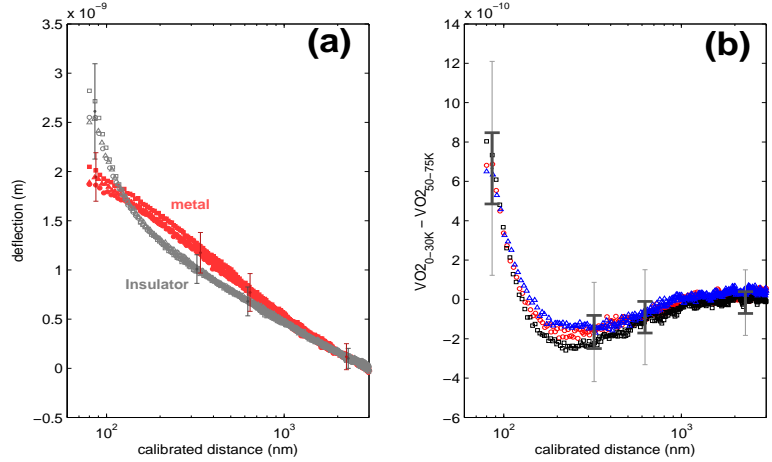

*Figure 2. (a) Three averages of RHT-distance curves (squares, circles and triangles) are shown for both the metallic phase and the insulating phase of VO2. Each average curve is an average of 9 measurements at different places. (Fig. 1 shows one set of raw data for one average). In (b) we show the resulting RHT difference curves between the metallic and insulating phase of $VO_2$. The three averages are shown with different colors. The thin error bars in (a) and (b) depict spatial variation of the measurements over the surface. The thick ones are the standard deviations of the average of 9 curves, which are a factor of 3 lower.*

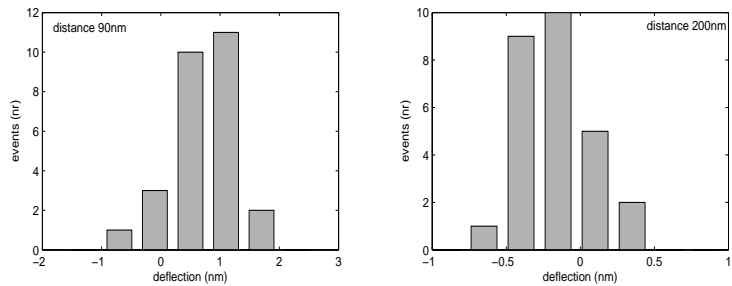

*Figure 3: Data distribution of all 27 measurements at different places on the surface for plate-sphere distances 90 and 200nm. Data is shown for the difference curve plotted in fig 2b, i.e. between the $\Delta T= 75K$ minus the $\Delta T=50K$ curve and the $\Delta T= 30K$ minus the $\Delta T=0K$ curve.*